      [1999/12/01 v1.4c Il Nuovo Cimento]
\documentclass{cimento}

\usepackage{graphicx}  
\title{A solution of the cusp problem in relaxed halos of dark matter}
\author{
E.~Mikheeva,\from{ins:x}\ETC\thanks{helen@asc.rssi.ru},
A.~Doroshkevich,\from{ins:x}, V.~Lukash\from{ins:x}}
\instlist{\inst{ins:x} Astro Space Centre of P.N. Lebedev Physics
Institute - Profsoyuznaya 84/32, Moscow 117997 Russia}
\PACSes{\PACSit{95.35.+d, 98.65.Dx, 98.80.-k}{}
\PACSit{---.---}{\ldots}}
\begin{document}

\maketitle

\begin{abstract}
We propose a solution of the cusp problem in framework of the
standard $\Lambda$CDM cosmology. To do this we describe the
linear and nonlinear periods of halo formation by the entropy
function of dark matter particles. This approach allows us to
take into account together the impact of both the processes of
nonlinear relaxation of compressed matter and the small scale
initial velocity perturbations in collapsed halos. We show that
such random velocities lead to the random variations of the
density profile of relaxed halos. As a rule, they suppress the
formation of cusp--like halos and favor the creation of
core--like ones. This approach allows us to reproduce
observed rotation curves, to explain their random scatter and
deviations from simulated ones.
\end{abstract}

\section{Introduction}
The cusp problem comes from numerical simulations predicted
the formation of divergent density profiles inside virialized
DM halos. However such a behavior of density profiles is not
seen in observations. This contradiction is known as "the cusp
problem".

Mathematically, this means that in the central part of
gravitationally-bounded halos we get the power-law density
profile
\begin{equation}
\rho(r)\propto r^{-\alpha},
\label{dprof}
\end{equation}
with $\alpha\ge 1$ in numerical simulations (e.g.~
\cite{ref:diemand}) and $\alpha< 1$ in observed galaxies
(e.g.~\cite{ref:deBlok, ref:swaters}). In the first case we
deal with cusp, in the second one -- with core.
In addition, the available observational data show a broad
scatter of $\alpha$ (see ~\cite{ref:deBlok, ref:swaters}) and
sometimes $\alpha>1$ are observed (for example, in rich
clusters of galaxies~\cite{ref:pointecouteau}). Majority of the
observed rotation curves of galaxies~\cite{ref:marchesini}
are well fitted by the Burkert's function with $\alpha=0$
(see~\cite{ref:burkert}) which differs significantly from the
simulated curves fitted by the Navarro-Frenk-White (NFW)
function with $\alpha=1$, (\cite{ref:nfw, ref:nfw2}).

This situation has created the problem discussed during many
years as an important problem of the standard $\Lambda$CDM
cosmology. More of that, there is a common myth that the
relaxation of collisionless DM particles always produces
density cusps.

\section{Physical nature of difference between cusps and cores}

To clarify the difference between core and cusp we
calculate pressure profile for DM halo. For power-law
density profile (\ref{dprof}) we get for  the mass of halo
\begin{equation}
M=M(r)=\int_0^\infty \rho(r)r^2\,dr\propto r^{(3-\alpha)}\,,
\end{equation}
and for the pressure, $p(r)$, within an equilibrium halo
\[
\frac{1}{\rho}\frac{dp}{dr}=-\frac{GM(r)}{r^2}\propto r^{(1-
\alpha)}\,,
\]
\begin{equation}
p(r)=C_1+C_2r^{2(1-\alpha)},
\end{equation}
where $C_1$ and $C_2$ are some constants. So, if $\alpha\leq 1$
then in the central part of halo the pressure is finite and we
have the core case. If $\alpha > 1$ then the pressure is infinite
what corresponds to the case of cusp. Evidently, $\alpha=1$ is
a critical value of $\alpha$ corresponding to the logarithmic
cusp formation.

The halo formation includes both processes of the reversible
matter compression and the irreversible relaxation. To
discriminate between them we will consider the entropy function,
$F(r)$. For the nonrelativistic DM particles with the isotropic
pressure it can be introduced as follows:
\begin{equation}
p(r)=\rho\langle v^2\rangle = nT = F(r) \,n^{5/3}(r)\,,
\end{equation}
where $v$ is one-dimensional random velocity, $\rho,\,n,\,\&\,T$
are the density, concentration and temperature of DM particles.
$F$ can be presented as a function of current radius of halo, $r$,
but it is more convenient to consider $F$ as a function of current
mass, i.e. mass inside of $r$. We call $F(M)$ by entropy mass
function:
\begin{equation}
F(M)\propto C_1 M^{\beta_1} + C_2M^{\beta_2}\propto M^\beta,
\end{equation}
\begin{equation}
\beta_1=\frac{5\alpha}{3(3-\alpha)},\quad
\beta_2=\frac{6-\alpha}{3(3-\alpha)},\quad
\beta\in(\beta_1,\beta_2)\,. \label{beta12}
\end{equation}
For the critical value of $\alpha=1$ we get
$\beta_{cr}= \beta_1 = \beta_2 = 5/6$.

The entropy mass function integrates the impact of irreversible
processes during all DM evolution and determines the profile of
relaxed halo. As is seen from (\ref{beta12}) for $\alpha < 3$ we
have $\beta_1\geq 0,\,\beta_2\geq 0$ and, so, $F(M)\rightarrow 0$
for $M\rightarrow 0$. However, the fraction of low entropy DM
particles in the central regions of halo in the cusp case is
larger than that in the case of core.

\section{Idea and method}

The discussions of the halo formation are usually concentrated
around the processes of DM relaxation which are main sources of
entropy. At the same time it is usually accepted that before
matter relaxation the DM entropy is negligible. However, in the
CDM model there are random correlated velocities down to very
small scales determined by the mass of DM particles. In a course
of the matter compression they are partly transformed to the
temperature of particles and, so, increase the entropy mass
function at small $M$. This effect can be estimated analytically.
Further on we will calculate the joint entropy in relaxed DM halos
summing the initial entropy given by small scale initial
perturbations and an entropy generated during non-linear
relaxation of collapsed matter.

Our further analysis includes the following steps:

\noindent 1. Instead of density-radius relations we use the
entropy-mass profiles.

\noindent 2. The mean initial entropy profile, $F_b(M)$, is
estimated as the entropy mass function of linear velocity
perturbations. Such identification assumes the almost adiabatic
matter compression and, so, partly underestimates the entropy
mass function $F_b(M)$ at small $M$.

\noindent 3. The entropy generated by the process of relaxation of
compressed matter, $F_r(M)$, is taken from results of N-body
simulations (e.g. NFW profiles).

\noindent 4. We combine the initial and generated entropies by
a simple way to set of joint entropy profiles and reconstruct
the circular velocities and resulting density profiles.

\section{Initial entropy mass function}

At the beginning we need to consider initial field of density
perturbations and its statistical properties.

Let us characterize the initial conditions in dark matter by three
gauge-invariant variables -- displacement of a matter point from
unperturbed position (the deviation from Hubble flows), $\vec
S(\vec x)$, the full velocity of matter, $\vec V(z,\vec x)$, and
comoving density perturbation, $\delta(z,\vec x)\equiv\rho(z,\vec
x)/\rho(z)-1$:
\begin{eqnletter}
\vec r(z,\vec x)& = & \frac{1}{1+z}\left[\vec x-g(z)\vec
S(\vec x) \right]\,,\\
\vec V(z,\vec x) & = & \dot{\vec r} = H(\vec r + g'\vec S)\,,
\end{eqnletter}
\[
\delta(z,\vec x) =  g(z)\cdot\delta(\vec x)\,,\quad\delta(\vec x)
= div\,\vec S = \partial S_i/\partial x_i\,,
\]
where the dot and prime mean derivative with respect
to the physical time and the function argument, $\vec r(z,\vec x)$
and $\vec x$ are the Eulerian (shear-free) and Lagrangian
(comoving) coordinates of a matter point.

In these terms perturbations related with the collapsed halo
as whole and conditional perturbations within the halo can be
separated as follows:
\begin{eqnletter}
\vec S(\vec x) & = & \vec S_{R}(\vec x) + \vec S_*(\vec x)\\
\delta(\vec x) & = & \delta_R (\vec x) + \delta_*(\vec x)
\end{eqnletter}
where quantities with index $R$ are related to a protohalo with
a linear size $R$, and quantities with index $*$ are related to
conditional perturbations which determine the corresponding
temperature and entropy of compressed matter,
\begin{equation}
\langle\vec S_*\rangle =\langle \delta_*\rangle =0\,\quad
\langle\vec S_*^2\rangle \equiv \sigma_*^2(R)=\int
P(k)\left[1-W(kR)\right]^2dk.
\end{equation} where $W(x)$ is a
usual spherical window function.

After some analytical and numerical calculations we obtain for
the power index $\beta$ of initial entropy mass function $F(M)$
for $M\leq 10^nM_\odot$ (see Table~\ref{tab:1}):
\begin{table}
  \caption{Power index of the function $F\propto M^\beta$
for halos with $M\leq 10^nM_\odot$.}
  \label{tab:1}
  \begin{tabular}{rcccc}
\hline
$n\equiv log (M/M_\odot)$ & 10 & 7 &  5  &  $\leq$ 5 \\
\hline
$\beta\equiv d\ln F(M)/d\ln M$ & 0.3 & 0.5 & 0.57 & 0.67\\
\hline
  \end{tabular}
\end{table}

It is easy to see that in all cases $\beta$ is less than its
critical value $\beta_{cr}=5/6$. This means that
initial entropy dominate in central regions of halo
what stimulates the formation of cores and suppresses the
cusp formation.

\section{Entropy generation during the matter relaxation}

The spherical collapse with the violent relaxation of compressed
matter have been  investigated in many papers. Thus, authors
of~\cite{ref:filgol} started from initial conditions with no
initial velocities and entropy
\begin{equation}
\delta M(r)=1-M(r)/\langle M(r)\rangle\propto \langle M(r)
\rangle^{-\varepsilon}\,,
\end{equation}
\begin{equation}
v_i=0,\quad
\delta\rho=\rho_{in}(r)-\langle\rho_{in}\rangle \propto
r^{-3\varepsilon}\quad \langle M(r)\rangle\propto r^3\,,
\end{equation}
and found that the density profile of relaxed matter, $\rho(r)$,
is approximated at small $r$ by power law:
\begin{equation}
\rho(r)\propto r^{-\alpha},\quad \alpha= \left\{
\begin{array}{cc}
2, & \varepsilon\leq 2/3\cr
9\varepsilon/\left(1+3\varepsilon\right), & \varepsilon\geq 2/3\cr
\end{array}.
\right.
\label{dns}
\end{equation}

The approach was extended in~\cite{ref:sikivi, ref:nusser}
where nonradial trajectories of DM particles were also considered.
The density profiles for relaxed halos was found to be similar to
previous one.

This problem have also considered in ~\cite{ref:gurzyb, ref:gz2}
for the initial conditions
\begin{equation}
v(r)=0,\quad \rho(x)=\rho_0(1-r^2/r_0^2)\,.
\end{equation}
Near the center of cloud the power law density profile was found
with
\begin{equation}
\rho(r)\propto r^{-\alpha},\quad \alpha\sim 1.7 -1.9\,.
\end{equation}
Anywhere initial velocities and entropy was accepted to be
equal to zero.

As it is seen from these relations, the model of
{\it violent relaxation} leads to the entropy distribution
\begin{equation}
M(r)\propto r,\quad F_g(r)\propto r^{4/3}\propto M^{4/3},\quad
\beta=4/3\geq \beta_{cr}\,,
\end{equation}
that implies the formation of cusp. However, in the central
regions of halos the entropy generated by this process becomes
negligible as compared with the initial entropy. Thus, the
influence of former one restricts the central density of
halos and prevents formation of cusp--like density profile.

The same conclusion is valid also for the simulated NFS --
density profile with $\alpha=\alpha_{cr}=1,\,\beta=\beta_{cr}=
5/6$ which represents the more general model of {\it hierarchical}
halo formation. The drop of these indices as compared with
previous ones indicates that in the case the initial entropy
is partly allowed for but owing to the small scale cutoff of
the simulated power spectra its impact is underestimated as
compared with estimates in Table 1.

\section{Rotation curves caused by the joint entropy}

The shape of rotation curves is determined by sum of initial and
generated entropy. To clarify influence both of them we model
the joint entropy mass function as follows:
\begin{equation}
F\left(M\right) = \sqrt{F_b^2+F_g^2},\quad F_b\equiv \kappa F.
\end{equation}
where the random factor $\kappa,\,0\leq\kappa\leq 1$, measures
the relative contribution of the initial entropy for different
halos.

Now we can calculate rotation curves and compare them with
observed and simulated ones. The results are presented in Fig.
1 where the circular velocities, $v_c$, are plotted versus
radius for six models of hierarchal ($\beta_g=5/6$, left plot)
and six models of violent ($\beta_g=4/3$, right plot) relaxation
processes for initial conditions with $\beta_b=0.333, 0.567,
0.667$ and for $\kappa\ll 1$ and $\kappa\simeq 1$. Both the
velocities and radius are normalized for their values in point
of velocity maximum, $v_{max}=v_c(r_{max})\geq v_c(r)$. The
curves are compared with the NFW and Burkert ones.

As is seen from this Figure, the rotation curves for our models
with $\beta_b=1/3$ and $0\leq\kappa_s\leq 1$ close the gap between
the Burkert and NFW rotation curves. So, we can expect that
properties of observed curves can be successfully reproduced by
our models with suitable parameters $\kappa$ and $\beta_b$. The
scatter of both initial and generated entropies can provides
required variations of these parameters and the shape of rotation
curves what, in turn, explains variations of observed rotation
curves.

Let us note also that for low mass halos with
$\beta_b=0.567$ and especially $\beta_b=2/3$ all rotation curves
with $0\leq\kappa\leq 1$ are concentrated nearby the NFW
profile. This fact indicates that probably for dwarf galaxies with
$M\leq 10^8M_\odot$ the observed rotation curves can be similar
to expected ones for the model of hierarchical clustering with
the NFW profile.

\begin{figure}
\includegraphics{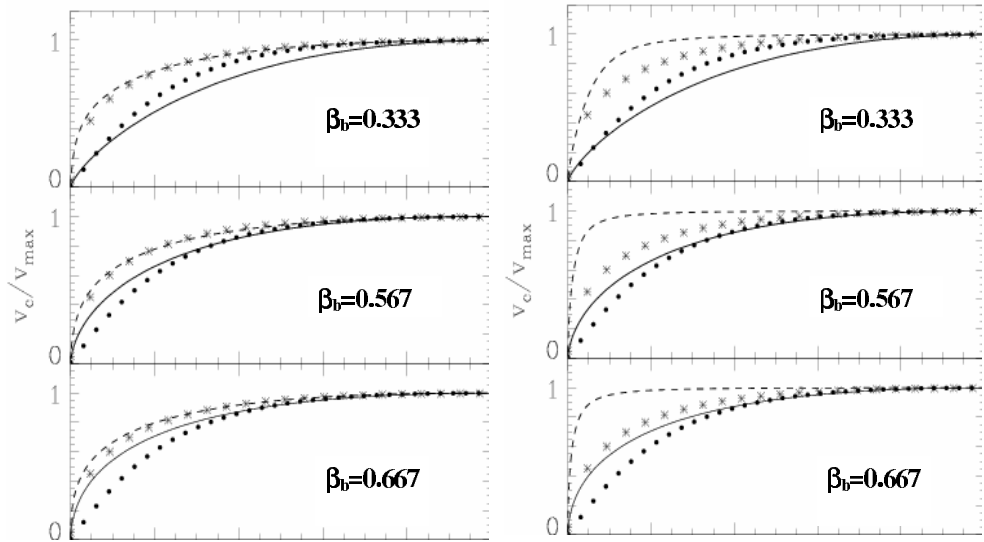}     
\caption{Normalized rotation curves are plotted for the models of
hierarchical (left side of the Figure) and violent (right side)
relaxation for $\kappa_s\ll 1$ and $\kappa_s\approx 1$ (dashed and
solid lines). NFW and Burkert fits are plotted by stars and dots,
respectively.}
\end{figure}

\section{Conclusions}

We conclude, that

1. The initial entropy can prevent the cusp formation for halos
with DM masses $10^8-10^{12}M_\odot$. For smaller and larger
galaxies and for clusters of galaxies the impact of the initial
entropy is attenuated.

2. The impact of the initial entropy allows to reproduce the
observed rotation curves and many helps to solve the so called
``cusp problem''.

\acknowledgments {Authors cordially thank Organizing Committee of
the Conference for hospitality. This work was supported by Russian
Foundation for Basic Research (grant numbers 05-02-16302 and
07-02-00886).}

\end{document}